\documentclass[webpdf,contemporary,large,numbers]{oup-authoring-template}%

\usepackage{url}
\usepackage{multirow}
\usepackage{array}

\usepackage{ulem}

\usepackage{booktabs}
\usepackage{color}
\usepackage{threeparttable}
\usepackage{graphicx}
\usepackage{natbib}

\theoremstyle{thmstyleone}%
\theoremstyle{thmstyletwo}%
\theoremstyle{thmstylethree}%

\begin{document}

\journaltitle{PNAS Nexus}
\DOI{DOI HERE}
\copyrightyear{2025}
\pubyear{2025}
\access{Advance Access Publication Date: Day Month Year}
\appnotes{Paper}

\firstpage{1}


\title[Social inequality and cultural factors impact the awareness and reaction during the cryptic transmission period of pandemic]{Social inequality and cultural factors impact the awareness and reaction during the cryptic transmission period of pandemic}

\author[a]{Zhuoren Jiang}
\author[b,$\ast$]{Xiaozhong Liu}
\author[c,d,f]{Yangyang Kang}
\author[c,d]{Changlong Sun}
\author[e]{Yong-Yeol Ahn}
\author[e]{Johan Bollen}

\authormark{Jiang et al.}

\address[a]{\orgdiv{School of Public Affairs}, \orgname{Zhejiang University}, \orgaddress{\street{866 Yuhangtang Rd}, \postcode{310058}, \state{Hangzhou, Zhejiang}, \country{P.R. China}}}
\address[b]{\orgdiv{Data Science}, \orgname{Worcester Polytechnic Institute}, \orgaddress{\street{100 Institute Road}, \postcode{01609}, \state{Worcester, MA}, \country{U.S.A.}}}
\address[c]{\orgname{Alibaba Group}, \orgaddress{\street{969 West Wen Yi Road}, \postcode{310030}, \state{Hangzhou, Zhejiang}, \country{P.R. China}}}
\address[d]{\orgdiv{College of Computer Science and Technology}, \orgname{Zhejiang University}, \orgaddress{\street{Yuquan Campus}, \postcode{310013}, \state{Hangzhou, Zhejiang}, \country{P.R. China}}}
\address[e]{\orgdiv{School of Informatics, Computing, and Engineering}, \orgname{Indiana University Bloomington}, \orgaddress{\street{107 S. Indiana Avenue}, \postcode{47405}, \state{Bloomington, IN}, \country{U.S.A.}}}
\address[f]{\orgdiv{Polytechnic Institute}, \orgname{Zhejiang University}, \orgaddress{\street{866 Yuhangtang Rd}, \postcode{310058}, \state{Hangzhou, Zhejiang}, \country{P.R. China}}}

\corresp[$\ast$]{To whom correspondence should be addressed: \href{email:email-id.com}{xliu14@wpi.edu}}

\received{Date}{0}{Year}
\accepted{Date}{0}{Year}


\abstract{The World Health Organization (WHO) declared the COVID-19 outbreak a Public Health Emergency of International Concern (PHEIC) on January 31, 2020. However, rumors of a ``\textit{mysterious virus}'' had already been circulating in China in December 2019, possibly preceding the first confirmed COVID-19 case. Understanding how awareness about an emerging pandemic spreads through society is vital not only for enhancing disease surveillance, but also for mitigating demand shocks and social inequities, such as shortages of personal protective equipment (PPE) and essential supplies. Here we leverage a massive e-commerce dataset comprising 150 billion online queries and purchase records from 94 million people to detect the traces of early awareness and public response during the cryptic transmission period of COVID-19. Our analysis focuses on identifying information gaps across different demographic cohorts, revealing significant social inequities and the role of cultural factors in shaping awareness diffusion and response behaviors. By modeling awareness diffusion in heterogeneous social networks and analyzing online shopping behavior, we uncover the evolving characteristics of vulnerable populations. Our findings expand the theoretical understanding of awareness spread and social inequality in the early stages of a pandemic, highlighting the critical importance of e-commerce data and social network data in effectively and timely addressing future pandemic challenges. We also provide actionable recommendations to better manage and mitigate dynamic social inequalities in public health crises.}

\boxedtext{This study comprehensively characterizes how emerging disease information spread through the Chinese population by examining large-scale chronological purchasing data of personal protective equipment on an eCommerce platform. Analyzing data from 94 million individuals, it reveals how socioeconomic status, social networks, geography, and cultural factors relate to awareness diffusion, uncovering substantial inequalities in information access and response. These findings expand the theoretical framework on awareness diffusion and social inequality during pandemics and highlight evolving characteristics of vulnerable populations, emphasizing early interventions tailored to specific communities. The study also provides practical recommendations for adopting flexible and dynamic public health strategies and offers new insights into the complex influence of cultural factors and social structures on pandemic responses.}

\maketitle

\section{Introduction}

The emergence of COVID-19 was accompanied by demand surges for PPEs (Personal Protective Equipments) and other consumer goods in nearly every country~\cite{burki2020global}. For instance, China experienced shortages of sanitizing wipes, face-masks, and other essential goods in early 2020~\cite{wu2020facemask,sodhi2023research}. Although official confirmation of the disease drastically accelerated demand, this demand surge began well before the official WHO announcement of a Public Health Emergency of International Concern (PHEIC), due to rumors of a ``\textit{mysterious virus}'' on social media that started circulating as early as December 2019~\cite{valentin2021monitoring,shi2022online}. This suggests that the information about an emerging pandemic can spread well before its official confirmation during its cryptic transmission period, and that the traces of such information may be detectable in rapid changes in demand for particular consumer goods. Although it has been several years since COVID-19 has become a pandemic, the relevance of examining this phenomenon persists, as we anticipate the next pandemic and socio-economic inequalities continue to shape how different demographics are affected by such shifts in demand and consumption, with vulnerabilities dynamically evolving based on the spread of awareness and information throughout the course of a pandemic. Hence, we examine behavioral data captured by a massive e-commerce platform during 88 days from December 1st, 2019 to February 26th, 2020, which is referred to as the ``cryptic transmission period''\footnote{During this time, all 94 million individuals in our dataset were observed to be fully aware of the pandemic. The WHO-China Joint Mission on COVID-19 reported the main findings of the outbreak on February 24th, 2020, which was two days before the end of our observation period.}.

To understand this phenomenon tying early awareness of a pandemic during its cryptic period to changes in consumer behavior and concomitant social inequities, we need to address two questions: (1) ``\textit{how does the awareness of an emerging pandemic spread through the various demographics of a population?}'' and (2) ``\textit{how does authoritative and officially vetted information from local, national, and international instances affect awareness diffusion?}'' Understanding the dynamics of pandemic awareness during the cryptic period can critically inform emerging disease surveillance, but it may also shed light on how demand surges and shortages of critical goods affect the public response to future health emergencies. It is particularly important to understand how the differential diffusion of awareness creates information gaps between different demographic cohorts, which can lead to socio-economic inequities that negatively affect a society's ability to respond to an emerging public health crisis involving an infectious disease~\cite{laajaj2022understanding,suh2022disparate,folayan2023multi}. In fact, in the case of the COVID-19 pandemic, shortages of PPEs and essential supplies, or the lack of equitable distribution, may have disproportionately affected disadvantaged population groups potentially contributing to higher infection rates and mortality~\cite{worby2020face,ye2022equitable}. In a public health emergency, enhancing public awareness and preparedness, and doing so in a socially and economically equitable manner, can be as important as mitigating the crisis itself~\cite{funk2009spread, kamalrathne2023need}. 

Here, we investigate the early spread of pandemic awareness in China by data mining a unique record of 150 billion query and purchase records for 94 million individuals from the largest eCommerce platform, Alibaba. This data was collected right at the crucial time period of the early COVID-19 outbreak (12/01/2019-02/26/2020), along with longitudinal records of purchasing history of the included individuals in the past 10 years, providing detailed geo-located shipping/gifting information from which 43.7 million families, 120.3 million schoolmates, and 25.8 million workmates relations were informed~\cite{cen2019trust}. By focusing on PPE-related queries and orders, this data enables investigations into the collective querying and purchasing behavior at the cryptic periods of the pandemic, leveraging changes in consumer demand as a proxy of awareness~\cite{addo2020covid,funk2009spread}. We examine the dynamic factors that are associated with the timing of the initial response for a variety of vulnerable populations in terms of their location, social relations, and demographics.

The eCommerce data recorded in this period provides a unique opportunity to study the evolving pandemic awareness throughout various socio-economic demographics for the following reasons. First, eCommerce is now the primary means of shopping for the Chinese population\footnote{https://www.statista.com/outlook/emo/ecommerce/china}. Second, eCommerce data collection reflects user behavior in its natural state. This non-intrusive approach offers an unbiased view of an individual's actual needs and preferences, mitigating social conformity and observer effects typically associated with traditional self-reported data. Third, unlike social media or search engine data, which often reflect only interactions and textual expressions, eCommerce data captures a substantial, real-time, and representative snapshot of actual purchasing behaviors, geographic distribution, and consumption patterns. It can enable comprehensive investigations into the associations between individual demographics and purchasing behaviors, which can be further enhanced by detecting longitudinal patterns. For instance, if an individual purchased pregnancy products 7 years ago, diapers 6 years ago, and toys recently, it can be inferred that the individual may have a child of age 7. 
This helps us analyze the connections between demographic factors, pandemic awareness, and public health risks. This allows the estimation of rich information to unravel associations between different demographic factors, pandemic awareness, and a variety of public health risks. Fourth, purchase records also contain gifting and shipping information, which enables multi-view social ties estimation. For example, when two people share the same home shipping address, we may infer a social or familial bond \cite{cen2019trust}. Furthermore, when shipping addresses pertain to the same company/school/dorm address, we can infer individuals' classmate/workmate relations \cite{cen2019trust}. This enables a detailed analysis of social networks and their impact on pandemic awareness and disaster planning. By leveraging these unique data characteristics, eCommerce data facilitates innovative research that can track the diffusion of pandemic awareness with high resolution across social, temporal, demographic, and geographical dimensions, at truly societal scales (a majority of the Chinese population)\footnote{More detailed dataset information, privacy and ethical considerations, and machine learning inference information are available in the Supplementary Information (SI).}. This data also allows for early detection of shifts in consumer behavior related to health concerns, providing valuable signals for timely public health interventions.

\section{Data Collection and Awareness Label Generation} 

We analyze how early awareness of COVID-19 spread through the various sectors, locations, social networks, and demographics of the Chinese population by leveraging a data set of 46.5 billion queries (randomly sampled) issued by 800 million individuals, covering 88 days from 12/1/2019 to 2/26/2020, a crucial period around the time WHO declared a public health emergency, here referred to as the ``cryptic transmission period." In addition, for the same population, we randomly sampled 150 billion historical queries and purchase behaviors, shipping and gifting addresses, etc. ranging from 2010 to 2019, to estimate individuals' demographic variables and their social relationships, i.e.,~family, workmate, and schoolmate networks (see SI (2.C)). 

Our reasons for selecting this time period and China as the focus of our research are as follows: \textbf{First}, China is the first country to report a large-scale outbreak of COVID-19~\cite{whoreprot2020}, with the initial confirmed cases reported in December 2019 in Wuhan, Hubei Province~\cite{huang2020clinical,li2020early}. As the first country to confront the pandemic, China's initial response provides invaluable insights into societal awareness and reactions during the crucial early stages of a global health crisis. \textbf{Second}, the chosen time period reflects distinctive social behavior responses. During this period, public knowledge of the virus was limited~\cite{wu2020new,li2020early}, and guidance from the government and healthcare institutions gradually became clearer~\cite{whoreprot2020}. Concurrently, the crucial resources were potentially being rapidly and unequally consumed~\cite{ye2022equitable,gozzi2023estimating}. Investigating public awareness during ``cryptic transmission period,'' characterized by incomplete information and high uncertainty, will help effectively identify vulnerable populations and understand how to enhance societal resilience in future crises, mitigating the potentially severe consequences of inequality. \textbf{Third}, China’s vast and complex social structure, characterized by wealth disparities, high population density, and significant population mobility, presents unique challenges. Studying how Chinese communities and individuals became aware and responded in this complex environment can provide valuable lessons for crisis management in similarly complex social settings. In conclusion, China's ``cryptic transmission period" during the COVID-19 pandemic offers a unique national context that can provide invaluable insights for global public health management and crisis response strategies.

We mark each individual $i$ at time $t$ as either ``aware'' ($L^{i}_t=1$) or ``not aware'' ($L^{i}_t=0$) following the definition of ``pandemic awareness'' proposed in \cite{funk2009spread}. $L^{i}_t$ is determined on the basis of whether they submitted any awareness-indicating query ($q^\star \in \mathbb{Q}$) three or more times at time $t$. We define a set of awareness-indicating queries $\mathbb{Q}$ that are deemed related to COVID-19 PPEs, i.e., ``\textit{(n95 or kn95 or kf94) \& Face mask}''. All queries in $\mathbb{Q}$ were found to discriminate well between pandemic-related and not-pandemic related search activity; the queries were nearly never issued before the start of the pandemic and experienced dramatic growth during the early pandemic, with negligible noise and an average increase (in search population) of 102,792\% compared to the same period from 2018 to 2019. Once $L^{i}_t=1$, individual's awareness status will not revert. Only individuals with active eCommerce behaviors (who made at least a purchase every month in the past 5 years) were employed for awareness modeling.  Following these filtering guidelines, we located 94,534,663 qualified individuals (11.8\% of all the 800 million individuals) for awareness analysis and model generation. By 02/26/2020, the last day of our observation period, all individuals' awareness labels had changed to 1. This study was reviewed by the IRB at Indiana University (protocol \#: 10521) and was determined not to constitute human subjects research, thus not requiring further review. The experiment and secured data processing methods were also reviewed and approved by the Alibaba legal department. All data used in this study have been anonymized and de-identified to prevent any possibility of identifying individuals. 

To provide exhaustive awareness analysis, following WHO guidance \cite{world2009pandemic}, we segment our COVID-19 cryptic transmission period into five sub-phases\footnote{Data from the normal phase can reflect background information. For instance, during the normal phase, individuals with children are more likely to search for masks compared to those without children.}, as shown in Table \ref{Tab:phase}.

\begin{table*}[htbp]
\centering
\small
\caption{Five Different Phases \cite{world2009pandemic} of the Cryptic Period of COVID-19 with Specific Real-World Events}
\label{Tab:phase}
\scriptsize
\begin{tabular}{l p{1.7cm} p{5cm} p{8.2cm}}
\textbf{Phase} & \textbf{Time Period} & \textbf{Definition} & \textbf{News/Social Events} \\
\midrule
\multirow{2}{*}{\textbf{Normal}} & 12/01/2019-12/30/2019 & All province-level awareness percentages $\leqslant$ 0.001\% & 12/08/2019: A COVID-19 case was reported in retrospective studies ~\cite{li2020early} \\
\midrule
\multirow{3}{*}{\textbf{Beginning}} & \multirow{3}{1.7cm}{12/31/2019-01/18/2020} & \multirow{3}{5cm}{First province-level awareness growth rate $>$ 100\% and national awareness percentage $>$ 0.001\%} & 12/31/2019: Wuhan MHC released a briefing (pneumonia outbreak)\\
 & & & 01/05/2020: Wuhan MHC reported 59 cases of viral pneumonia\\
 &  &  & 01/16/2020: Strict exit screening measures activated in Wuhan\\
\midrule
\multirow{3}{*}{\textbf{Growth}} &  \multirow{3}{1.7cm}{01/19/2020-01/22/2020} & Second province-level awareness growth rate $>$ 100\% and national awareness percentage $>$ 0.001\% & \multirow{3}{*}{01/20/2020: China NHC confirmed human-to-human transmission}\\
\midrule
\multirow{3}{*}{\textbf{Peak}} & \multirow{3}{1.7cm}{01/23/2020-01/26/2020} & \multirow{3}{5cm}{More than 95\% provinces' awareness growth rates $>$ 10\% and national awareness percentage $>$ 0.1\%} & 01/23/2020: Wuhan lockdown\\
 &  & & 01/24/2020: Hubei activated first-level public health emergency\\
 &  & & 01/25/2020: China activated first-level public health emergency\\
\midrule
\multirow{3}{*}{\textbf{Post-Peak}} & \multirow{3}{1.7cm}{01/27/2020 - 02/26/2020} & \multirow{3}{5cm}{More than 95\% provinces' awareness growth rates $<$ 10\%  and continuously drop for at least 3 days\\} & 
01/31/2020: WHO declared the novel coronavirus outbreak a PHEIC\\
& & & 02/02/2020: Wuhan launched quarantine strategies\\
& & & 02/11/2020: WHO named COVID-19 officially \\ 
\bottomrule
\end{tabular}
\end{table*}

\begin{figure*}[htbp]
\centering
  \includegraphics[width=0.9\linewidth]{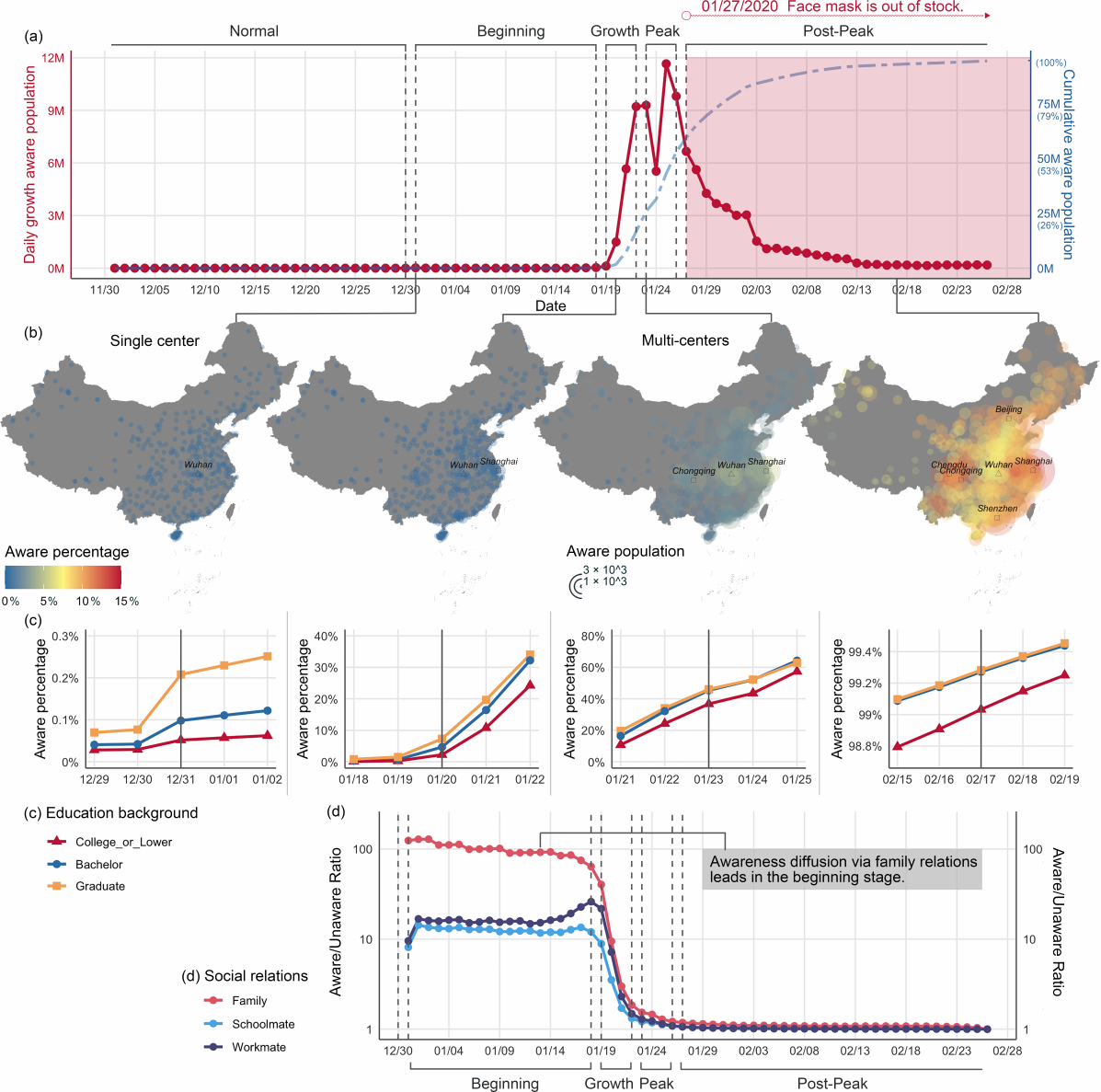}
  \caption{The Patterns of Awareness Diffusion (5 Phases) (a) The diffusion of awareness and reaction. The red Y-axis on the left represents the daily growth in the aware population (red dotted line), and the blue Y-axis on the right corresponds to the daily cumulative aware populations (blue dash-dotted line). There are two peaks on the daily trends: 9,289,545 newly aware on 01/23/2020 (Wuhan lockdown) and 11,655,320 newly aware on 01/25/2020 (30 provincial-level regions activated first-level public health emergency). (b) The geographic awareness distributions (366 cities) on four representative days of different phases. The size of the circle indicates aware population, the color indicates the awareness percentage, the triangle represents the epicenter, and the squares represent the city with most aware individuals. The initial awareness surged from the epicenter Wuhan (in the beginning phase), and gradually spread across the whole country with the increasing in pandemic severity. (c) The awareness percentage trends of different education groups (for four representative days of different phases). In the beginning phase, the group with graduate degrees led a higher aware percentage after Wuhan MHC released a pneumonia outbreak briefing (12/31/2019). The similar trends can be observed in growth phase when China NHC confirmed human-to-human transmission (01/20/2020). During the peak phase (after Wuhan lockdown), the aware percentages of graduate and bachelor groups were close and higher than the college or lower group. This trend continued during the post-peak phase. (d) Neighborhood awareness ratio (between aware individuals' aware neighbor percentage and unaware individuals' aware neighbor percentage) following three types of social relations. In the beginning phase, all three social relations can diffuse pandemic information efficiently (ratios greater than 8), while family relation showed the highest diffusion efficiency (ratios $\in \left [ 63.9, 128.7 \right ]$). After the growth phase, it is hard to differentiate aware and unaware individuals' neighbors (ratio converges to 1). The basemap used in Figure 1 is sourced from the National Platform for Common Geospatial Information Service, China (Map Approval Number: GS (2024) 0650) and complies with Chinese laws and regulations. This map is for illustrative purposes only, used to visualize research data without any political or territorial assertions.}
  \label{fig:oges}
\end{figure*}

\section{Societal Spread of COVID-19 Awareness}

Figure \ref{fig:oges} (a) illustrates daily (red) and cumulative (blue) growth of the number of individuals whose status changed to ``aware''. The trend exhibits rapid growth between Jan. 19 and 26 (after ``\textit{National Health Commission of China confirmed human-to-human transmission}''), and experiences two peaks on Jan. 23 (``\textit{Wuhan lockdown}'') and Jan. 25 (``\textit{China activated first-level public health emergency}''). Note that since Jan. 27, most PPEs (e.g., face masks) were out of stock in China, implying that individuals who changed status to ``aware'' afterwards might not have been able to obtain PPE.

\subsection{Geography, Education \& Social Relation Analysis} 

Figure \ref{fig:oges} (b) visualizes the geographic distribution of the aware population and the awareness percentages of the populations of 366 major cities in Mainland China over time. During the normal and beginning phases (see Table \ref{Tab:phase}), the epicenter Wuhan had the most aware individuals. The number of aware individuals in Shanghai (a leading metropolis) increased significantly in the growth phase. At the peak and post-peak phases, the awareness percentage also increased rapidly in other major cities. Geographically, awareness was initially correlated with the severity of the pandemic, but subsequently spread across the entire country, while the population's response seemed to be most pronounced in large cities. 

Figure \ref{fig:oges} (c) shows awareness levels across educational backgrounds. In the beginning phase, the postgraduate group (master or PhD) exhibited the highest awareness percentage which continued into the growth phase, revealing that educational background could be an important variable driving early pandemic awareness or preparedness. Highly educated demographics seem to have had an advantage in picking up on news and rumors of the epidemic effectively, resulting in earlier and therefore more extensive access to scarce resources when they were still available. Throughout the peak and post-peak phases, individuals with postgraduate and bachelor degrees share similar awareness patterns. Individuals with lower educational levels exhibit the lowest awareness percentage across all phases (e.g., on average, the awareness percentage of the postgraduate group is 4.39 times that of the college or lower group during the beginning phase). As a result, they are likely to have responded more slowly to the emerging pandemic, and may have been more vulnerable in terms of pandemic preparedness and access to PPE.

In Figure \ref{fig:oges}(d), we explore changes in how awareness propagates through different types of social relations, by defining a ``social neighborhood awareness ratio'', i.e., the ratio of the aware neighbor percentage of aware individuals to that of unaware individuals across different phases (see SI (3.D) for the detailed math definition). In this context, the term "social neighbors" refers to individuals who are directly connected to a person through a particular social relationship. We infer familial, work, and educational relations (represented by the \textit{family}, \textit{workmate}, and \textit{schoolmate} networks respectively) in this study. For each social relation type, the neighborhood awareness ratio reveals the degree to which awareness diffuses through the specific type of relation. From Fig. \ref{fig:oges} (d), we observe that, in the beginning phase, familial relations can be particularly influential for information diffusion, and work relations can be more important than educational relations. In the later stages, the potential impact of all these relations may have declined rapidly due to the aggressive speed with which the pandemic spread across society. 

Changes in awareness across the different social relationship types can be informative. Note that one cannot use this data to infer causality, due to the (latent) homophily~\cite{shalizi2011homophily}. We caution against overinterpreting these differences, as each type of social tie may be influenced by varying levels of homophily and other confounding variables.

\begin{figure*}[htbp]
\centering
  \includegraphics[width=0.9\linewidth]{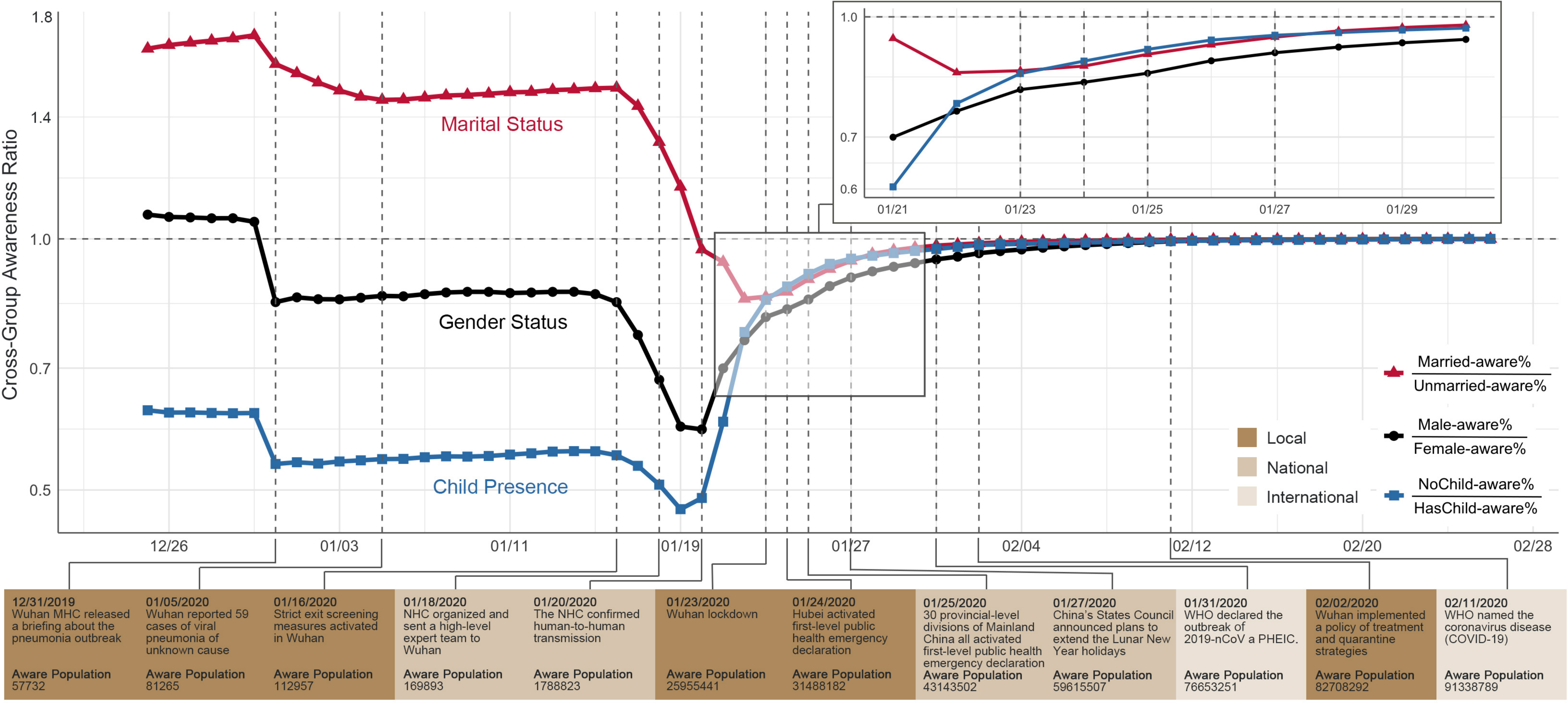}
  \caption{The Awareness Trends across Different Demographic Groups with Important News Events. Each line represents cross-group awareness ratio ($R=P_{G_{1}}/P_{G_{2}}$. $R$ is the cross-group awareness ratio, and $P_{G_{i}}$ is the percentage of aware people in the group $G_{i}$). When the first official pandemic briefing released (12/31/2019), females, with-children, and unmarried groups reacted more quickly (cross-group awareness ratio trend-lines dropped). After strict screening tests were activated in Wuhan (01/16/2020), females, with-children, and unmarried groups showed stronger awareness strengths (awareness ratio trend-lines dropped and kept declining). After the NHC confirmed human-to-human transmission (01/20/2020), male and without-children groups began to show significant awareness strengths (awareness ratio trend-lines began to rise). It was not until the Wuhan lockdown (01/23/2020) that the married group began to show a relatively stronger level of awareness compared to the unmarried group (married-aware/unmarried-aware ratio trend-line began to rise). A base-10 log scale is applied for the Y axis.} 
  \label{fig:gcm}
\end{figure*}

\begin{figure*}[htbp]
\centering
  \includegraphics[width=0.9\linewidth]{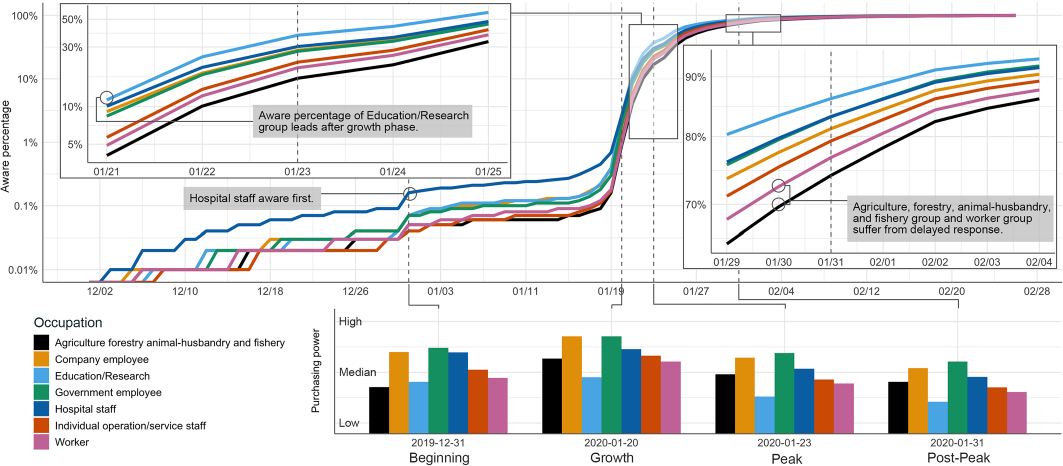}
  \caption{Awareness (percentage) Patterns for Different Occupation Groups (upper sub-plot); a base-10 log scale is applied for the Y-axis; the left cut-out zooms in on details for 5 days around the Wuhan lockdown (01/23/2020); the right cut-out zooms in on details for 7 days around WHO declared the new coronavirus outbreak (01/31/2020) in the post-peak phase. The hospital staff kept the highest awareness percentage (0.16\%-0.45\%) in the whole beginning phase. In the growth phase, the education/research group surpassed hospital staff and became the most aware group (0.39\%-25\%), while agriculture forestry animal-husbandry and fishery were the least group (0.16\%-10.09\%). The peak phase showed a similar pattern; education/research was the most aware group (37.24\%-66.25\%) while agriculture forestry animal-husbandry and fishery were the least one (16.86\%-44.03\%). The gaps of different occupation groups shrank during the post-peak phase. The lower sub-plot visualizes four representative days of different phases, and the Y-axis is the average purchasing power of the aware population from different occupation groups. Results show that high-income people respond to the emerging pandemic more quickly than low-income people.}
  \label{fig:oap}
\end{figure*} 

 \subsection{Gender, Marital and Child-presence Analysis} 
 
 Figure~\ref{fig:gcm} shows differences between early awareness changes according to marital, gender, and child-presence status. The cut-out zooms in on details for 10 days after the NHC (National Health Commission of China) confirmed human-to-human transmission. The Male-aware/Female-aware ratio trend line shows how awareness changed over time according to gender. Notably, the ratio does not change monotonically, indicating that certain events can trigger differential effects in male and female populations. For instance, the two events when (1) ``\textit{Wuhan MHC (Wuhan Municipal Health Commission) released a briefing about the pneumonia outbreak}'' (12/31/2019) and (2) when ``\textit{strict exit screening measures were activated in Wuhan}'' (01/16/2020) corresponded to a drop in the Male-aware/Female-aware ratio (from 1.05 to 0.6), whereas the event of ``\textit{The NHC confirmed human-to-human transmission}'' (01/20/2020) corresponded to a sharp increase in the ratio. This phenomenon suggests that women may be more sensitive or receptive to indications of an emerging pandemic. This gender gap however seems to be narrowing as the severity of the pandemic and public awareness increases, echoing prior studies \cite{bord1997gender,stockemer2021covid,otterbring2022pandemic} which show gender differences in risk perception and risk aversion.

 NoChild-aware/HasChild-aware ratio trajectory shows that individuals with children are much more likely to react to the emerging pandemic than those without children. This is observed throughout the entire data coverage (ratio $\leq$ 1), with the ratio reaching 0.48 before ``\textit{The NHC confirmed human-to-human transmission}''. This ratio trend dropped 13\% when ``\textit{Wuhan MHC released a briefing about the pneumonia outbreak}'' (12/31/2019), indicating that individuals with children have greater awareness of an emerging pandemic during the beginning phase of cryptic transmission period. 
 
The trend of Married-aware/Unmarried-aware ratio reveals that a higher percentage of individuals in the married group would search PPEs, well before ``\textit{Wuhan MHC released a briefing on the pneumonia outbreak in the city}'' (peak on 12/30 with ratio = 1.73). As pandemic concerns diffused through the entire population, the growth rate of awareness in the unmarried group accelerated. However, it was not until ``\textit{Wuhan lockdown}'' (01/23/2020) that the ratio began to rise again, indicating a resurgence in the relative awareness growth of the married group compared to the unmarried group.

\subsection*{Occupation \& Purchasing Power Analysis}

Occupation and purchasing power (indicating income) inferred from eCommerce data reveal economic and social inequality during pandemic early stages.

As shown in Figure \ref{fig:oap}, the awareness percentage of hospital staff leads other occupations during the normal and beginning phases (0.16\%-0.45\%). When ``\textit{Wuhan MHC released a briefing on the pneumonia outbreak}'' (on 12/31/2019), hospital staff became aware more rapidly compared to other communities. The awareness percentage of the research and education group which includes teachers, researchers, and students (2.67\%) surpassed that of hospital staff group (2.65\%) on 01/20/2020, when ``\textit{China NHC confirmed human-to-human transmission}''. Throughout the growth phase (01/19/2020-01/22/2020), the research and education group (0.39\%-25\%), hospital staff (0.69\%-20.6\%), white-collar employees (0.39\%-18.5\%), and government employees (0.3\%-18.02\%) maintained a high awareness percentage, whereas the awareness of ``agriculture, forestry, animal husbandry, and fishery'' group (0.16\%-10.9\%), blue-collar workers (0.18\%-12.19\%), and individual operation and service staff (0.17\%-13.74\%) grew more slowly. The peak phase (01/23/2020-01/26/2020) shared a similar pattern on 01/26/2020, after ``\textit{all 30 provincial-level regions activated first-level public health emergency}''. The awareness percentage of research and education (66.25\%), hospital staff (58.53\%), government employees (56.89\%), white-collar employees (56.13\%), and individual operation and service staff (52.38\%) exceeded 50\%, but the awareness percentages of agriculture, forestry, animal husbandry and fishery (44.03\%) and blue-collar workers (48.38\%) grew more slowly. It was not until ``\textit{WHO officially named COVID-19}'' on 02/11/2020 that the awareness percentages of all occupation groups exceeded 95\%. 
We also estimated individual purchasing power as a proxy to their income levels, enabling a comparison of awareness levels across income levels, which is crucial to gauge socio-economic inequities. For all occupations, high-income groups seemed to have become aware of pandemic more quickly than low-income ones. We visualize the purchasing power of aware individuals on four representative days from each phase in Figure \ref{fig:oap} which highlights the income gap between communities that became aware earlier and later. Based on daily awareness growth rates, we found that the highest purchasing power groups (levels 6 and 7) maintained high growth across the beginning, growth, and peak phases. In contrast, the lowest purchasing power group (level 1) was slower to be aware, only becoming the fastest-growing in the later stages (after 01/25/2020). Among occupations, ``hospital staff'' and ``education/research'' showed rapid awareness growth in the earlier phases; while ``workers'' and ``agriculture, forestry, animal husbandry, and fishery'' were slower to be aware, only becoming the fastest-growing groups after 01/27/2020. When comparing the impact of occupation and purchasing power on awareness, occupational groups generally showed higher daily maximum growth rates, leading on 64 out of 88 days. This suggests that occupation may have a stronger influence on awareness than purchasing power. Detailed information on awareness percentages and daily awareness growth rates is available in SI (2.D).

\begin{figure*}[htbp]
\centering
  \includegraphics[width=\linewidth]{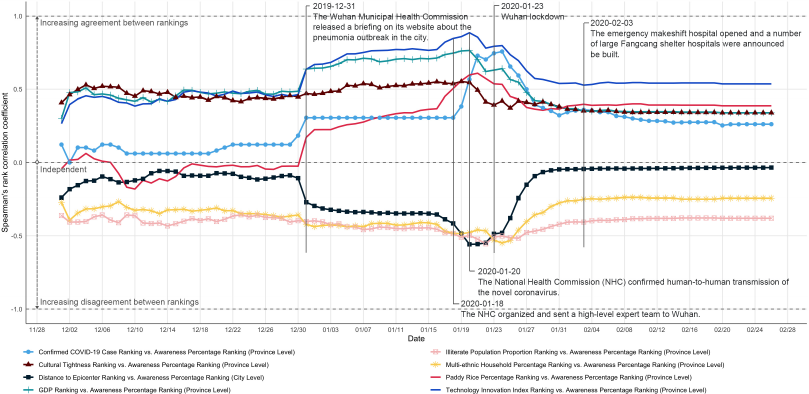}
  \caption{The Trends of Geography-related Spearman's Rank Correlation Coefficients: Distance to Epicenter (Wuhan) vs. Awareness Percentage (for 366 major cities); Confirmed COVID-19 Cases vs. Awareness Percentages, GDPs vs. Awareness Percentages, Cultural Tightness vs. Awareness Percentages, Paddy Rice Percentages vs. Awareness Percentages, Technology Innovation Indexes vs. Awareness Percentages, Illiterate Population Proportions vs. Awareness Percentages, Multi-ethnic Household Percentages vs. Awareness Percentages (31 provinces). Following the beginning phase, factors such as the distance from the epicenter, the proportion of the illiterate population, and the percentage of multi-ethnic households exhibit a negative correlation with awareness. In contrast, the number of confirmed COVID-19 cases, GDP, percentage of paddy rice, and technological innovation index show a positive correlation with awareness.}
  \label{fig:correlation}
\end{figure*}

Overall, in the beginning stage of the pandemic, we found that high-income, well-educated individuals, and females became aware earlier than other groups, potentially affording them better opportunities to access and purchase scarce resources. Meanwhile, social relationships, familial in particular, may have played a key role in the diffusion of awareness.

\subsection{Geographic Socio-economic, Cultural, and Structural Factor Analysis} In terms of the geographic distribution of awareness dynamics (Figure \ref{fig:oges} (b)), we found that awareness is strongly related to the geographic location. This observation inspired us to further explore the geographic factor(s) that shaped pandemic awareness, and their dynamics. We first focus on three main geographical factors: \textit{distance to the pandemic epicenter (Wuhan)}, \textit{confirmed COVID-19 case numbers}, and \textit{local Gross Domestic Product (GDP)}. We rank the geographic locations based on the mentioned three factors and compare these to their ranking based on population awareness percentages for each day in the time period under consideration. See SI (4.C) for detailed information.

We measured the statistical dependence between factor-based geo-rankings and awareness percentage geo-rankings by calculating their Spearman rank-order correlation~\cite{pirie1988spearman}. Figure \ref{fig:correlation} visualizes the trends of these three rank correlation coefficients. Overall, a city's awareness percentage and its distance to the epicenter are negatively correlated, and a province's confirmed COVID-19 cases and GDP (province rankings) are positively correlated with awareness. The degree of positive correlation between cases and awareness began to increase on 12/31/2020 (0.31) and rose rapidly from 01/19/2020 (0.38), one day after ``\textit{the NHC organized and sent a high-level expert team to Wuhan}''. It peaked on 01/24/2020 (0.76), when ``\textit{Hubei activated the first-level public health emergency}'', one day after the ``\textit{Wuhan lockdown}''. Subsequently, the coefficient decreased from 0.76 to 0.26. The correlation coefficient between GDP and awareness percentage exhibited a similar trend, where the degree of positive correlation started to grow notably on 12/31 (0.64), reached peak on 01/20/2020 (0.76), and then fell back (0.76 to 0.34). In other words, economic gaps matter in COVID-19 awareness, and people from lower-income locations are less aware of the pandemic than those from higher-income locations. This information gap can lead to inequalities in pandemic preparedness.

Additionally, we examined the time-evolving correlations between a series of socio-cultural and structural factors and regional awareness distributions. These factors include \textit{cultural tightness}~\cite{gelfand2011differences,chua2019mapping}, \textit{percentage of paddy rice~\cite{talhelm2020historically}}, \textit{technology innovation index}~\cite{China_Innovation_Report_2020}, \textit{proportion of illiterate population (aged 15 and above)}, and \textit{percentage of multi-ethnic households}. We found significant positive and negative correlations between these factors and awareness during the cryptic transmission period, which became more pronounced as the pandemic progressed: these correlations began to strengthen notably during the beginning phase, peaked during the growth or peak phase, and weakened in the subsequent stages. Specifically, cultural tightness was positively correlated with awareness. Higher illiteracy rates and a greater proportion of multi-ethnic households were associated with lower awareness. In contrast, higher paddy rice percentages and a higher technological innovation index were positively correlated with awareness. See SI (4.C) for detailed information.

These findings can be explained by cultural dimensions theory~\cite{hofstede1980culture}, tightness-looseness theory~\cite{gelfand2011differences}, and social capital theory~\cite{hawe2000social}. These theories highlight how inequalities in social structure, cultural diversity and resource distribution can lead to differences in information dissemination and processing, behavioral responses, and adaptability. Cultural dimensions theory suggests that cultural tightness, characterized by strict norms, can facilitate more effective information dissemination and quicker responses during pandemics~\cite{mattison2021digital}, helping to unify public health responses and reduce information gaps. The connection between rice farming and tight social norms~\cite{talhelm2020historically} may explain why regions with greater paddy rice cultivation exhibit higher awareness during severe pandemic phases. Historically, the cooperative capabilities developed through rice farming foster greater collective mobilization and social cohesion in response to public health threats, consistent with tightness-looseness theory. Social capital theory emphasizes the negative impact of social inequality on pandemic responses. High illiteracy rates often indicate an unequal distribution of educational resources~\cite{street2011literacy}, while language barriers and cultural differences in multi-ethnic households can hinder information dissemination~\cite{clark2020disproportionate}. Enhanced technological capabilities may improve risk perception~\cite{lima2005risk} and scientific prevention behaviors~\cite{qin2024mass}, leading to more effective awareness. Overall, these insights suggest that in collectivist cultures with unequal social capital, pandemic awareness depends on the level of available social capital. Groups with higher social capital achieve greater awareness, while those with lower social capital face disadvantages. Cultural tightness, emphasizing strict norms and collective action, can help mitigate these disparities by promoting consistent and timely awareness. However, rising social inequalities weaken overall awareness, especially for disadvantaged groups.

\subsection{The Impact of Real-world Events on Population Awareness} Our results hint at factors that are associated with awareness and may shape a population's reaction to an emerging pandemic. During the cryptic transmission period, several communities can advance a faster awareness response to local and public events, e.g., local news from the epicenter. When the pandemic news spread nationally, awareness spread to all communities. Local, national, and international news can make different contributions to the diffusion of awareness. In general, the impact of national events outperforms local ones, e.g., \textit{``China NHC confirmed human-to-human transmission''} (01/20/2020) is the landmark of the growth phase, while some local events alerted certain communities. Meanwhile, different awareness hysteresis effects may result from various events, e.g., after \textit{``Wuhan MHC released a briefing about the pneumonia outbreak''}, the aware population increased by 10\% within 1.5 hours, 50\% after 58 hours, and 100\% after 9 days. Contrastingly, after \textit{``China NHC confirmed human-to-human transmission''}, the aware population increased by 10\% just in 38 seconds, 50\% after 4 hours and 34 minutes, and 100\% after 12 hours and 26 minutes.

\section{Modeling the Dynamics of Public Awareness}\label{sec:model}

\subsection{Dynamic Awareness Modeling with Regression} 
\begin{figure*}[htbp]
\centering
  \includegraphics[width=0.9 \linewidth]{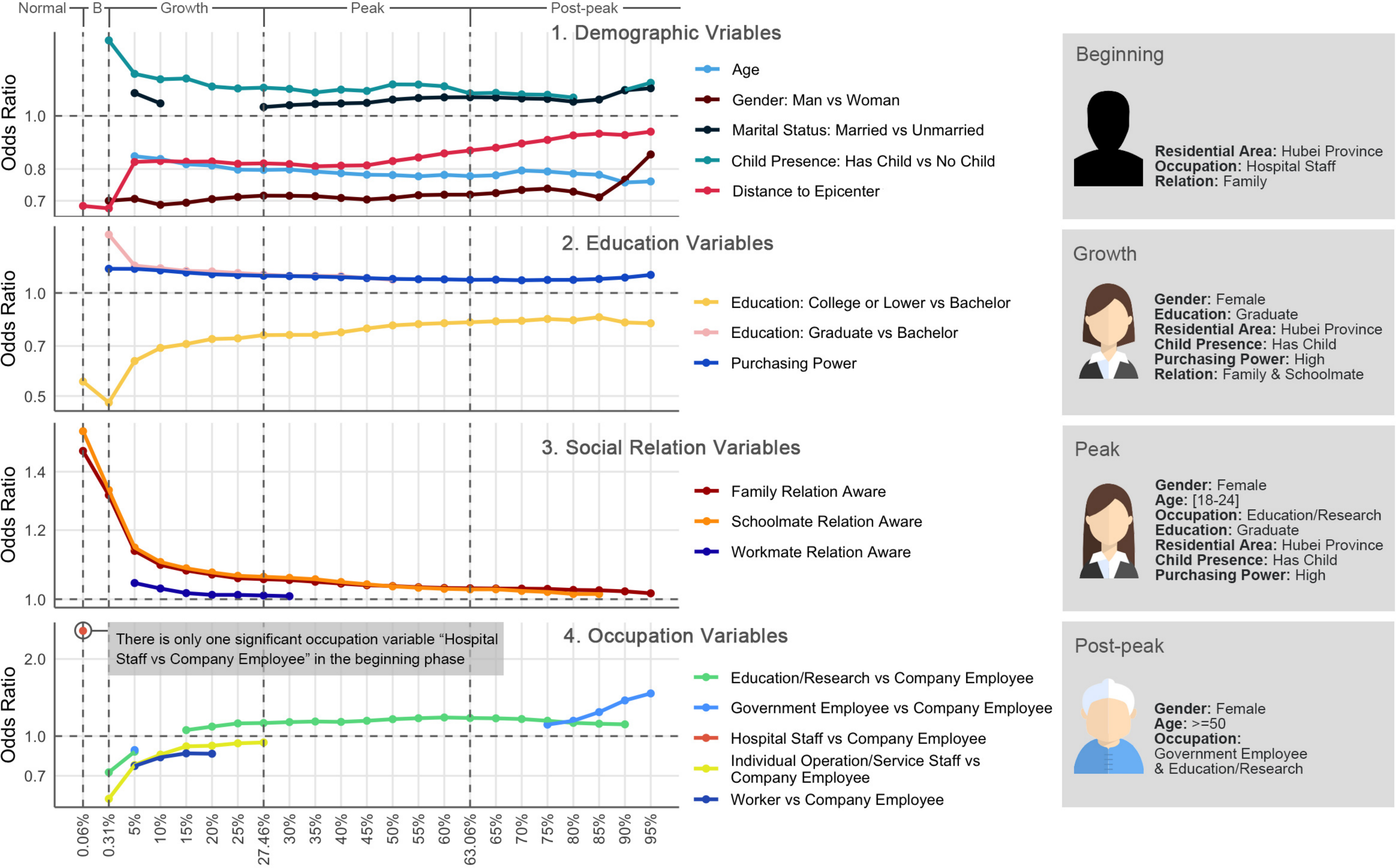}
  \caption{Logistic Regression Models (left), at different awareness percentage points, visualize the time-evolving trends of odds ratios of demographic and social relation features. X-axis is aware percentages and Y-axis is the odds ratios of variables. The typical characteristics of an aware individual (right) across the different phases.}
  \label{fig:regreesionprofile}
\end{figure*}

Our study focused on a descriptive analysis of the socio-economic factors that may affect an individual's ability to make adequate preparations for their health and protection. Can pandemic awareness be meaningfully predicted from these factors? We investigate this matter by modeling individual pandemic awareness with a logistic regression model \cite{mcdonald2009handbook} where the dependent variable is an individual's Boolean awareness label set $\{L^{i}_t\}$ (individual $i$'s awareness label, ``\textit{aware}'' or ``\textit{not aware}'', at time $t$). We employed 11 different independent variables to predict an individual's awareness status: \textit{Gender, Age, Occupation, Education, Distance to epicenter}, \textit{Purchasing power, Child presence, Marital status}, and \textit{Family/Schoolmate/Workmate awareness percentages} (see Figure \ref{fig:regreesionprofile}). More detailed feature descriptions can be found in SI Table S3. Since we can condition this awareness model for each time $t$, it allows us to characterize an individual's evolving awareness status over time.

For the experiment, we randomly sampled 100,000 individuals (from 94,534,663 individuals) to train time-evolving logistic regression models. The remaining individuals are retained to estimate the heterogeneous social networks required for some of the independent variables (familial, educational, and professional relationships). We fit a logistic regression model for each time $t$ when the overall aware percentage increased by 1\% (from 1\% to 95\%) plus the time points when important real-word events occur, yielding 106 models. Each logistic regression model estimates the log-odds of an individual becoming aware of the pandemic, i.e., the outcome that $L^{i}_t = 1$. The results are depicted in Figure \ref{fig:regreesionprofile}. For each timestamp, only significant variables ($p<0.05$) are presented, and Y-axis represents the odds ratio (see SI (5.B) for the detailed regression models and results).

The following observations can be made from the outcomes of these models. \textbf{First}, the geo-location, gender, child-presence, and age are the significant characteristics of an individual in the response to the emerging pandemic. Individuals close to the epicenter are able to respond effectively ($\mathrm{OR}_{distance}<1$, $p<0.05$). The odds of being aware would significantly decrease for older individuals ($\mathrm{OR}_{age}<1$, $p<0.05$) during the growth to post-peak phases. \textbf{Second}, highly educated and wealthy individuals respond to the emerging pandemic effectively. In the beginning phase (i.e., awareness percentage = 0.06\%), when PPEs and other resources were still widely available, compared to the individuals with bachelor degree, the low-educated individuals (college or lower) are at a disadvantage ($\mathrm{OR}=0.55$, $p<0.05$) to become aware of the pandemic, and this disadvantage for them continues throughout the whole observation period. Compared to the individuals with bachelor degree, individuals with Ph.D or master degrees show advantages in the response to the pandemic ($\mathrm{OR}>1$, $p<0.05$) from the growth to mid-peak phases. The individuals with high purchasing power are able to respond effectively ($\mathrm{OR}>1$, $p<0.05$) in the whole observation period. \textbf{Third}, compared to white-collar company employee, hospital staff can be aware quickly ($\mathrm{OR}=2.58$, $p<0.05$) in the beginning phase; and the individuals with other occupations are at a disadvantage ($\mathrm{OR}<1$, $p<0.05$) to become aware in the growth phase. \textbf{Fourth}, the family and schoolmate relations may be the critical relations for the awareness diffusion, especially in the early stages. In the beginning phase, family ($\mathrm{OR}_{family}=1.48$, $p<0.001$) and schoolmate ($\mathrm{OR}_{schoolmate}=1.56$, $p<0.001$) relations are both significant.

With respect to variables related to chronological ``Typical Aware Individual'', by leveraging significant features ($p < 0.05$), we characterize the typical individual for each phase who is aware of the emerging pandemic. In the beginning phase, the aware individuals can be close to the epicenter, work in a hospital, and have aware family member(s). In the growth phase, aware individuals are more likely to be well-educated, female, has-child, close to epicenter, and high income. The aware family and schoolmate relations can be the vital factors to enhance their awareness probability. In the peak phase, well educated, young (age $\in$ [18,24]), female individuals with ``\textit{education/research}'' occupation are more likely to be aware. In the post-peak phase, a typical aware individual can be a senior (age $\geqslant$ 50) female individual. The visualized typical aware individual can be found in Figure \ref{fig:regreesionprofile}.

\section{Discussion and Conclusion}
\subsection{Data and Methodological Strengths}
\textbf{Early Time Coverage:} this study offers a comprehensive analysis of the cryptic transmission period of COVID-19, from December 1, 2019, to February 26, 2020, capturing the earliest public reactions to the pandemic. Unlike previous research that primarily focuses on the post-outbreak stage~\cite{lu2021collectivism,chung2021socioeconomic,berger2024inequality}, this study provides valuable early evidence for understanding the formation of awareness and the dynamics of social behavior during the initial phase of the pandemic. \textbf{Large-Scale Data Analysis:} utilizing data from 94 million individuals and 150 billion query and purchase records from an e-commerce platform, it employs a large-scale data analysis to reveal behavioral differences among various socioeconomic groups, offering a deeper understanding of the dynamic shifts in social inequality during a public health crisis. The data scale in this study far surpasses that of previous studies~\cite{kalocsanyiova2023inequalities}. \textbf{Data Authenticity and Reliability:} by leveraging e-commerce behavior data in its natural state, the study avoids self-report biases and observer effects. Compared to studies that rely on surveys or self-report data~\cite{lu2021collectivism,chung2021socioeconomic,english2022historical,damette2023face}, this approach can more accurately reflect actual decision-making behaviors of individuals. \textbf{Interdisciplinary and Innovative Approach:} it introduces an innovative interdisciplinary approach by integrating demographic characteristics with heterogeneous social networks and applying deep learning techniques to large-scale behavioral data, offering a new perspective for analyzing social dynamics and advancing data-driven methods in social science research.

\subsection{Theoretical Contributions}
\textbf{First}, this study highlights the dynamic nature of the pandemic awareness among different socioeconomic groups by analyzing mask-related search behaviors rather than actual mask-wearing. We demonstrate that the spread of pandemic awareness not only evolves over time but also can be impacted by social inequalities. This finding broadens previous research on COVID-19  inequalities~\cite{10.1371/journal.pone.0254114,clark2020disproportionate,altmejd2023inequality,kontokosta2024socio,yerkes2024gender} and pandemic awareness~\cite{funk2009spread,despres2020communication}, emphasizing that public awareness is shaped by a combination of socioeconomic background, social capital distribution, and the progression of the pandemic. It enriches the theoretical understanding of how awareness and vulnerabilities shift with changing conditions, knowledge diffusion, and socio-economic factors. \textbf{Second}, this study finds that populations in regions with stricter social norms and greater cultural tightness exhibit higher levels of awareness during severe phases of a pandemic's cryptic transmission period. Previous research~\cite{lu2021collectivism,kitayama2022culture} has shown a positive relationship between collectivism and pandemic compliance. Our research further reveals a more nuanced relationship, showing that cultural tightness and tight social norms may not only influence compliance but also enhance awareness under conditions of uncertainty, thereby fostering more effective collective mobilization and social cohesion during public health threats. These findings extend the application of cultural dimensions theory~\cite{hofstede1980culture} and tightness-looseness theory~\cite{gelfand2011differences}, highlighting the complex and dynamic influence of cultural factors on pandemic awareness. \textbf{Third}, by examining the specific roles of family, schoolmate, and workmate relations in the spread of awareness, this study provides new insights into how relational dynamics function in the context of a pandemic. Our findings indicate that during early pandemic stages, family and schoolmate relations were identified as significant channels for awareness diffusion. This finding challenges the traditional emphasis on broader community or institutional-level channels, suggesting that personal and direct relationships may be more effective at certain stages of pandemic response. This offers a more detailed understanding of how awareness is distributed and diffused across different social network structures.

\subsection{Practical Implications}
\textbf{Utilizing E-Commerce Data for Dynamic Strategy Adjustment:} this study suggests that analyzing behavior data on e-commerce platforms during the early stages of a pandemic can provide real-time insights into the spread of awareness. Public health strategies should use this data to dynamically adjust their focus to increase pandemic awareness and promote preventive behaviors. \textbf{Targeted Interventions for Subdivision of Disadvantaged Communities:} public health policies should prioritize interventions tailored to disadvantaged communities, ensuring equitable access to information and resources. Deliberate attention must be given to subdivision of disadvantaged communities during each stage of a pandemic's emergence. By addressing socio-economic risk factors and improving information and resource access for groups with lower social capital, policymakers can bridge awareness gaps and reduce disparities in pandemic outcomes, enhancing community resilience. \textbf{Optimizing Communication Through Social Networks:} public health authorities should effectively utilize social networks for information dissemination in the early stages of a pandemic and continuously adapt their strategies as the situation evolves. Early interventions can prioritize leveraging family and schoolmate networks to ensure rapid information dissemination to socially disadvantaged groups, thereby raising awareness and promoting preventive behaviors more effectively. As the pandemic evolves, these strategies can expand to broader community engagement, ensuring that measures remain adaptive and responsive to changing conditions. \textbf{Adapting Public Health Strategies to Cultural Contexts:} the study's findings indicate that different cultures and social structures lead to varying responses and dynamics of awareness spread during a pandemic. Public health policies should adapt to these cultural differences across various pandemic phases to enhance effectiveness. For instance, in regions characterized by tight social norms, public health authorities may need to bolster external validation and disseminate information through trusted sources to mitigate potential delays in awareness. Additionally, leveraging cultural tightness to enhance pandemic awareness and preparedness, such as designing public health campaigns to reinforce existing norms of compliance and collective action, may improve the speed and consistency of responses.

\subsection{Conclusion}
Our analysis reveals how the information about COVID-19 outbreak unfolded in China, showing a conspicuous information gap between different populations, in particular during the early stages of the pandemic. This information gap, combined with subsequent shortages of essential PPE and other supplies, could induce inequities in the health outcomes of millions. Online purchase patterns show that the most vulnerable demographics are also the last to become aware of an emerging health crisis, and therefore least favorably placed to take precautions to protect themselves, their families, and their wider social network. This study leverages a large eCommerce dataset to investigate individual search and purchase characteristics at the individual, familial, institutional, and societal social level, observing the traces of emerging awareness through a population and its (inequitable) diffusion. This study not only expands the theoretical framework concerning awareness diffusion and social inequality during pandemics but also offers practical recommendations for public health policy. It emphasizes the importance of adopting flexible and dynamic response strategies during public health crises and provides new insights and practical guidance on understanding the complex influence of cultural factors and social structures on pandemic responses.

\section{Acknowledgments}
The authors thank the editor and anonymous reviewers, Zekun Wang for their valuable help and feedback. 

\section{Supplementary Material}
Supplementary material is available at PNAS Nexus online.

\section{Funding}
This work is supported in part by funds from the National Science Foundation (NSF: \# 2122232 and \# 1636636), National Natural Science Foundation of China (NSFC: \# 72104212), and Defense Advanced Research Projects Agency (DARPA: HR001121C0168).

\section{Author contributions statement}
X.L. and Z.J. designed research; Z.J., X.L., and Y. K. performed research; Z.J., Y. K., and C.S. analyzed data; X.L., J.B., Y.Y.A., and Z.J. wrote the paper.

\section{Competing Interest Statement}
The authors declare no competing interest.

\section{Previous presentation}
These results were never previously presented.

\section{Preprints}
There is no preprint of this article published.

\section{Data availability}
Due to privacy concerns and restrictions imposed by the data usage agreements, the raw data cannot be made publicly available. To support the understanding of our findings, we provide comprehensive statistical data, along with the 106 logistic regression results and related code files in publicly accessible repositories on Figshare. Daily growth rates and cumulative percentages of awareness and feature variables for 94,534,663 individuals (Dec 1, 2019 – Feb 26, 2020) are available at \url{https://doi.org/10.6084/m9.figshare.26831710} and \url{https://doi.org/10.6084/m9.figshare.26832400}. Feature distributions for 100,000 randomly selected individuals at 106 critical time points are available as follows: for aware individuals at \url{https://doi.org/10.6084/m9.figshare.26863462}, for unaware individuals at \url{https://doi.org/10.6084/m9.figshare.26863477}. The overall feature distribution for the 100,000-individual sample is at \url{https://doi.org/10.6084/m9.figshare.26863618}. Details of the 106 regression models are at \url{https://doi.org/10.6084/m9.figshare.24131109}, and related code is at \url{https://doi.org/10.6084/m9.figshare.24131130}. An anonymized, de-identified, and desensitized sample dataset is available upon compliance with data management regulations and reasonable request, which should be directed to the corresponding author, Xiaozhong Liu at \url{xliu14@wpi.edu}.

\bibliographystyle{plain}
\bibliography{reference}

\end{document}